\documentclass[12pt]{article}
\usepackage{epsfig,rotate,wasysym}
\usepackage{natbib}
\usepackage{xcolor}
\usepackage{ulem}
\usepackage{framed}

\usepackage{titlesec}
\titlespacing*{\section}{0.0\baselineskip}{0.5\baselineskip}{0.5\baselineskip}

\usepackage[disable]{todonotes}
\usepackage{pdfpages}
\title{Cryogenic cometary sample return}

\usepackage{times}  

\begin{document}

\newcommand\ajwsection[1]{\ \\ \noindent\underline{\bf{#1}}}
\newcommand\ajwsubsection[1]{\ \\ \noindent\underline{#1}}
\newcommand\ajwbfsubsection[1]{\ \\ \noindent\underline{\bf{#1}}}

\newcommand\bra[1]{{<#1|}}

\newcommand\ket[1]{{|#1>}}

\newcommand\zack[1]{{\textcolor{blue}{#1}}}
\newcommand\zackdel[1]{{\textcolor{red}{\sout{#1}}}}
\renewcommand{\em}{\itshape}

\def\window{Si$_3$N$_4$}
 
\newif\ifincludefigures
\newif\ifnofigures

\def\etal{{\em {\em {\em et al.}}}}

\includefigurestrue

\pagestyle{myheadings}
\markright{Cryogenic cometary sample return}

\thispagestyle{empty}



\addtocounter{page}{-1}
\pagenumbering{roman}
\pagenumbering{arabic}


\addtocounter{page}{-1}

\clearpage

\noindent {\Large \bf Cryogenic Comet Sample Return} \ \\ \ \\

\noindent {\large Andrew J. Westphal$^1$, Larry R. Nittler$^2$, Rhonda Stroud$^3$,  Michael E. Zolensky$^4$, Nancy L. Chabot$^5$, Neil Dello Russo$^5$, Jamie E. Elsila$^6$,  Scott A. Sandford$^7$, Daniel P. Glavin$^6$, Michael E. Evans$^4$,  Joseph A. Nuth$^6$,   Jessica Sunshine$^8$, Ronald J. Vervack Jr.$^5$, Harold A. Weaver$^5$ } \\
$^1$Space Sciences Laboratory, U. C.  Berkeley, $^2$Carnegie Institution of Washington, $^3$Naval Research Laboratory, $^4$NASA Johnson Space Center, $^5$Johns Hopkins University Applied Physics Laboratory $^6$NASA Goddard Space Flight Center, $^7$NASA Ames Research Center,$^8$University of Maryland  \\

\clearpage

\noindent {\Large \bf Cryogenic Comet Sample Return} \ \\ \ \\

\noindent {\large Andrew J. Westphal$^1$, Larry R. Nittler$^2$, Rhonda Stroud$^3$,  Michael E. Zolensky$^4$, Nancy L. Chabot$^5$, Neil Dello Russo$^5$, Jamie E. Elsila$^6$,  Scott A. Sandford$^7$, Daniel P. Glavin$^6$, Michael E. Evans$^4$,  Joseph A. Nuth$^6$,   Jessica Sunshine$^8$, Ronald J. Vervack Jr.$^5$, Harold A. Weaver$^5$ } \\
$^1$Space Sciences Laboratory, U. C.  Berkeley, $^2$Carnegie Institution of Washington, $^3$Naval Research Laboratory, $^4$NASA Johnson Space Center, $^5$Johns Hopkins University Applied Physics Laboratory $^6$NASA Goddard Space Flight Center, $^7$NASA Ames Research Center,$^8$University of Maryland  \\

\vspace*{-0.3in}

\section{Executive Summary} 
\label{section:summary}

Comets are unequalled repositories of the raw materials from which our solar system was made.  They likely formed in the outer regions of the protosolar nebula where they incorporated primitive presolar materials, volatiles resident in the outer disk, and more refractory materials from throughout the disk.  Since formation, the primordial materials in comet nuclei  have been stored at very low temperatures and protected by an overburden of surface materials, and are thus likely preserved in a largely unaltered state.

The return of a sample of volatiles (i.e., ices and entrained gases), along with other components of a cometary nucleus, will yield numerous major scientific opportunities.  Such samples will provide an unparalleled look at the primordial gases and ices present in the early solar nebula, enabling insights into the gas phase and gas-grain chemistry of the nebula.  Isotopic abundances in the volatiles can be used to better understand presolar and protosolar chemistry and address the importance of comets for the delivery of Earth's water and other volatiles.  Direct measurements of cometary volatiles will shed light on the processes that lead to cometary activity (dust emission, tail formation, fissioning, etc.).  Full understanding of the volatiles in comets will provide key information for interpreting the vast amount of telescopic data obtained from cometary comae, and this will place the entire returned sample in context with the greater cometary database.  Finally, understanding the nature of ices in their microscopic, petrographic relationship to the refractory components of the cometary sample will allow for the study of those relationships and interactions and a study of evolutionary processes on small icy bodies.

The previous 2013-2022 Decadal Survey included a study of a Flagship-class cryogenic comet nucleus sample return mission, given the scientific importance of such a mission. However, the mission was not recommended for flight in the last Decadal Survey, in part because of the immaturity of critical technologies.  Now, a decade later, the scientific importance of the mission remains and relevant technological advances have been made in both cryo instrumentation for flight and laboratory applications.   Such a mission should be undertaken in the next decade.  {\bf It is therefore urgent to conduct a new mission concept study of a cometary volatile sample return mission now, to properly inform the upcoming Decadal Survey discussions and development.}

\pagebreak
\setcounter{page}{2}

\begin{framed}
\noindent A cryogenic comet return mission would require: \\
\ \ \ $\bullet$ Further studies of thermal requirements necessary for preservation of petrological contexts \\
\ \ \ $\bullet$ A trade study of mass and power requirements for cryogenic systems \\
\ \ \ $\bullet$ Development of sampling technology that maximizes the return of primitive ices \\
\ \ \ $\bullet$ A trade study on mission architecture for two possible return modes \\
\ \ \ $\bullet$ Continued investment in cryogenic sample curation \\
\ \ \ $\bullet$ Continued investment in cryogenic sample handling and analysis techniques
\end{framed}

\begin{figure}[h!]
 \vspace*{-0.0in}
\hspace*{1.2in}\includegraphics[width=4.0in]{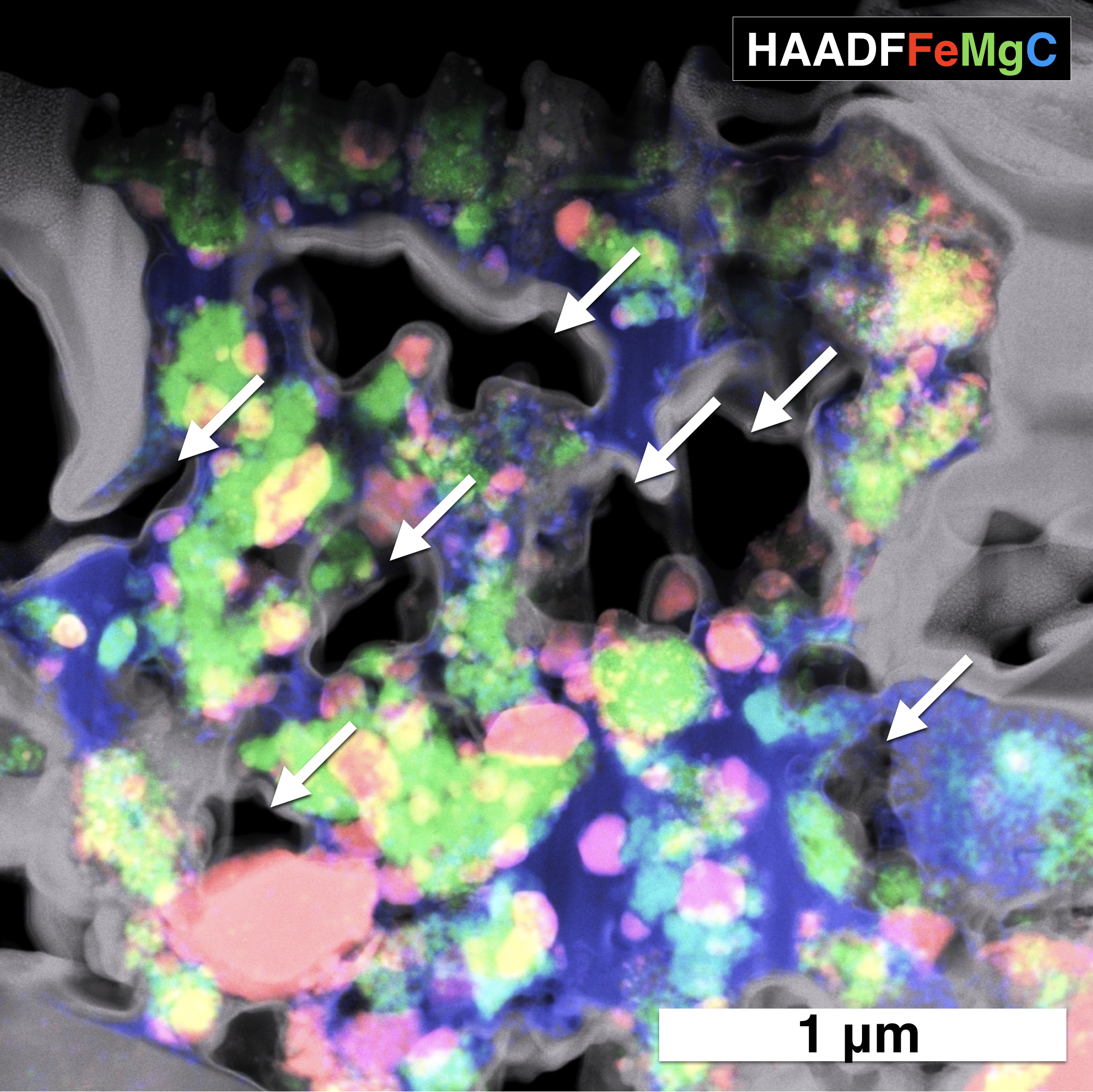}
 \vspace*{-0in}
 \caption{Transmission Electron Microscope (TEM) elemental map of a thin section of an interplanetary dust particle of likely cometary origin.  A heterogeneous assortment of nanoscale glassy and crystalline materials makes up this complex aggregate.  Voids, indicated by arrows, were probably filled with ices that did not survive. They beg the question:  {\em What are we missing?} (Image courtesy of Zack Gainsforth) }
\vspace*{-0.1in}
\label{fig:idp}
\end{figure}

\section{Terra Incognita:  cryogenic extraterrestrial materials} 
\label{sec:intro}

Comets preserve the most primitive building blocks of the Solar System. Having formed in the protosolar disk outside the volatile snow lines, comets are rich in ices \citep{bockeleemorvan2005}, although the origins of these ices (interstellar, protostellar, and/or presolar disk) are still unclear. The dust component of comets includes both high-temperature (formed in the inner disk) and low-temperature phases \citep{Brownlee2006}, so the ices are also likely to reflect a range of formation and processing histories \citep{Mousis2016}. Such ices are the building blocks of giant planets and their atmospheres. Understanding their formation may inform our understanding of exoplanet formation \citep{Cieza2016}. The simple molecules in these ices may undergo reactions to form more complex organics, creating the potential for a rich inventory of prebiotic organics \citep{Bernstein1995}.  The delivery of volatiles and organics to the prebiotic Earth may also have played a key role in making it habitable and in the origin of life \citep{Oro1961,ChybaSagan1992,Sandford2016}. The unique combination of low-temperature, high-temperature, organic, and inorganic phases makes comets superior time capsules of the earliest stages of Solar System history.

Just as refractory cometary materials are complex on a nanometer scale  \citep{Brownlee2006} and require analysis with laboratory instruments with sub-nm spatial resolution (Fig. 1), the icy components of comets are likely similarly complex (Fig. 2). Thus, a full understanding of the volatiles requires the same kind of coordinated, high-spatial resolution analysis that has been so productive for the study of cometary mineral components, but at cryogenic temperatures. These techniques are continuously improving, with a behavior reminiscent of Moore's Law \citep{westphal2016}. The range of micro- and nano-analytical instrumentation available in the laboratory today -- including electron, optical, infrared, x-ray, atom, and ion microprobes -- enables chemical, mineralogical, petrological, and isotopic analyses of refractory materials approaching or achieving atomic resolution.  No less important are increasingly sophisticated sample preparation techniques that enable these analyses.  These analytical and sample preparation techniques will be even more spectacular by the time of a cryogenic sample return in the early 2040s.

\begin{figure}[ht]
 \vspace*{-0.0in}
\hspace*{1.2in}\includegraphics[width=4in]{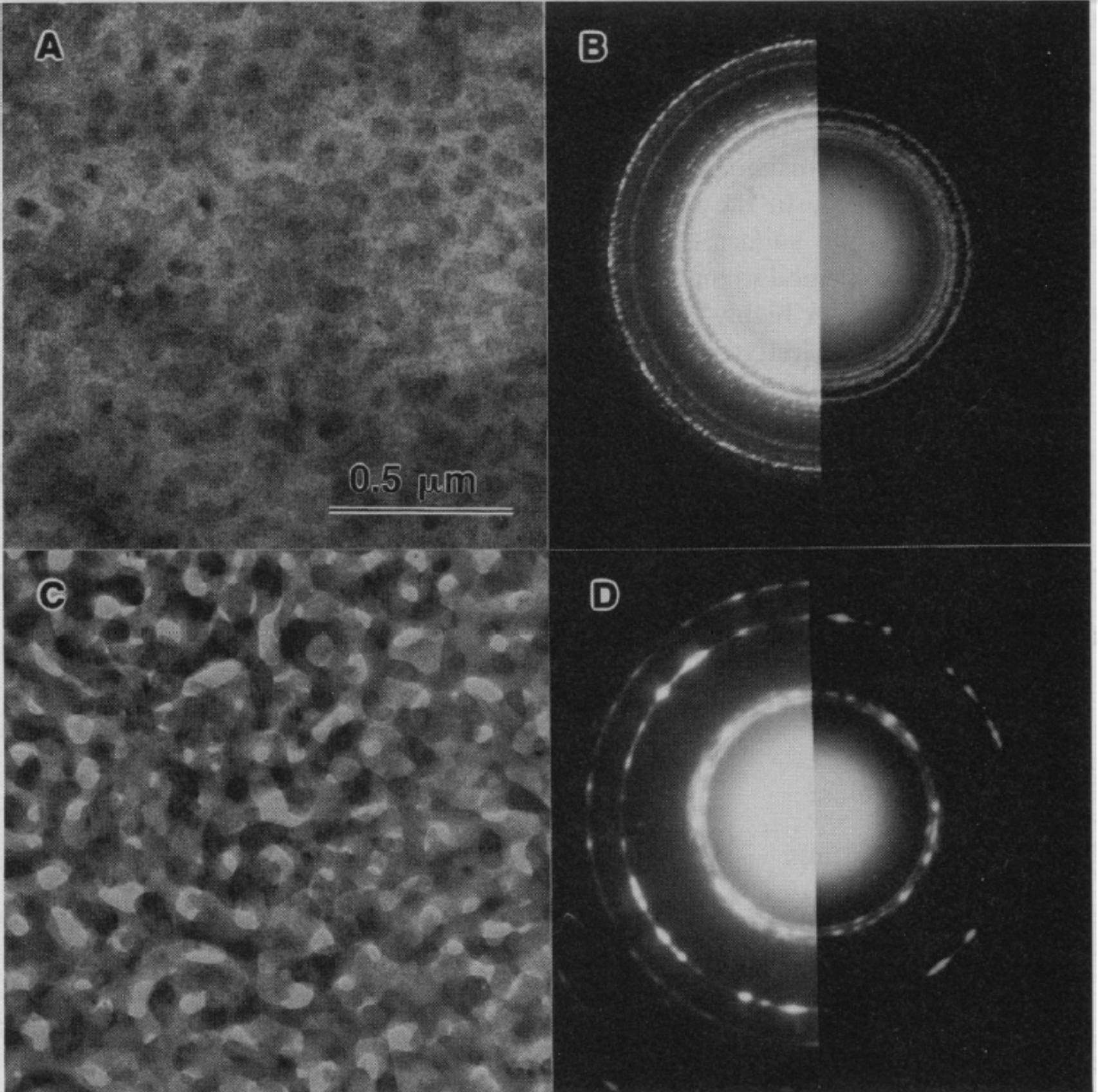}
 \vspace*{-0in}
 \caption{TEM study of a H$_2$O:methanol (2:1) ice, showing the effect of  warming through 120\,K (A,B) to 150\,K (C,D).  As the ice warms it becomes first a clathrate hydrate of methanol and then transforms into a highly porous hexagonal H$_2$O ice.  See \protect\citet{blake1991} for details. Preservation of clathrates and ice phases is one factor that will drive the sample storage temperature requirement for a cryogenic comet sample return mission.}

\label{fig:icetem}
\end{figure}


No natural astrophysical ices have ever been studied in the laboratory, although the structures, chemistry, and behaviors of astrophysical ice analogs have received considerable attention \citep{SandfordAllamandola1988}.  These laboratory studies give us an indication of the rich scientific insights that can be gained from the study of actual astrophysical ices.

We are unaccustomed to thinking of ices through a mineralogical/petrological lens, but at cryogenic temperatures, ices can be regarded as mineral components of rocky material like any other.  This is truly {\em Terra Incognita}, as a sample from a natural cryogenic (10s of  K) environment is unprecedented in any setting; currently, we can only make educated guesses about the nature of these materials on a microscopic scale.   The high-priority science questions to be addressed by coordinated mineralogical, petrological, chemical and isotopic analyses of cometary materials from a {\em cryogenic sample} in which the ices are preserved are thus truly fundamental, including: 

\noindent
{$\bullet$} What are the principal molecular components of the ices?  What structures can be recognized, and what are their sizes?  Are the ices homogeneous or heterogeneous? \\
{$\bullet$} How are the volatile components, {some newly discovered by Rosetta in comet 67P/C-G \ \citep{Rubin2019,Drozdovskaya2019},}  (CO$_2$, CO, O$_2$, NH$_3$, CH$_4$, HCN, HCNO, CH$_3$NO, CH$_3$CN, HC$_3$N,  {at least 23 other simple species and their isomers; complex organics such as glycine; ammonium salts \citep{Altwegg2020};} noble gases; etc.) distributed on small scales within cometary materials?  Are they trapped in more ``refractory'' ices, or do they exist as distinct phases? \\
{$\bullet$} What ice phases are present, and how are they distributed?  Are the ices amorphous or crystalline?  If amorphous, is the ice in the low-density or high-density form?  If crystalline, is the ice in a cubic or hexagonal structure or in the form of a clathrate? If clathrates are present, what form are they in?  Are Type II methanol-containing clathrates present? \\
{$\bullet$} How are the various volatile ices and organics distributed?  What is their spatial relationship with each other and with minerals, etc.?   \\
{$\bullet$} Do the D/H, $^{15}$N/$^{14}$N, $^{17}$O/$^{16}$O, $^{18}$O/$^{16}$O, and $^{13}$C/$^{12}$C isotopic ratios in the ices and volatile organics vary with molecular carrier and on what size scales? \\
{$\bullet$} Are there icy analogs of presolar grains, that is, presolar condensates with enormous isotopic anomalies? \\
{$\bullet$} How do the composition and physical structure of the ices drive and influence cometary activity and long-term evolution of the comet parent body? 

Curation and handling techniques are on target to be ready to accept cryogenic extraterrestrial samples by the anticipated conclusion of a successful cryogenic mission in the 2040s. Analyses that can address these questions are not currently fully available but are already being enabled by current Department of Energy (DOE) investments in cryogenic sample handling and analysis. 



{
\section{Sampling requirements}

To maximize the return of primitive ices, the sample acquisition system will acquire material from well below the upper processed layers of the nucleus,  preserving stratigraphy from sampling through curation. The thickness of the processed layers is poorly constrained and probably variable.  Even for the ice-coated grains in the regolith we will obtain valuable information on the structure of the ice in modern comets such as the existence of amorphous ice, clathrates, etc., that have the potential to power outbursts and alter the chemistry of the surface and coma.   Sample site reconnaissance will be required to balance safety and ease of sampling with the need to sample a site with high likelihood of containing primitive ices.
}

\section{Preservation of ice petrology}

{Retention of many volatile species within their microscopic context would  require  maximum sampling,  storage,  transport and curation temperatures significantly below those at which the original samples were collected. We suggest storage at or below 60\,K \citet{veverka2008} as that allows use of a solid Ar phase change medium during recovery and is below the temperature at which samples are likely to be sampled.}
At higher temperatures, these species might be retained within the sample, but, depending on their structure, could have been sufficiently mobile that their original petrological context would have been lost.  For example, amorphous H$_2$O ices undergo an amorphous-amorphous phase transition at $\sim$80 K that allows locate rearrangement of the ice matrix and allows some volatile species trapped in the ice to escape \citep{SandfordAllamandola1988}.  Highly volatile species (e.g., CO, O$_2$, noble gases, etc.) would be lost from the ice phase if not trapped in more abundant H$_2$O-rich ices.   { Long-duration, reliable storage at 60\,K is practically achievable with current technology (section \ref{section:cryocooler}).} 

\section{Cryocooler technology}
\label{section:cryocooler}

\textbf{ Long-duration, robust cryocooler technology now has extensive heritage on spacecraft.  This is a game-changer for cryogenic sample return.}
An 80\,K Stirling cycle cryocooler was first deployed for spacecraft use in 1991 on the UARS ISAMS mission. Since this initial flight, cryocoolers have been flown on more than sixteen other spacecraft and several high altitude balloon missions such as COSI and GRIPS. The RHESSI cryocooler operated for more than 15 years. This experience base is primarily in the 50\,K to 150\,K temperature range. Current missions near launch and in development are setting the groundwork for 5\,K to 20\,K operational temperatures. This mission experience makes cryocoolers a key technology for consideration in the cryogenic system. 

Cryogenic systems for comet sample return will be driven by three key mission phases: the sample extraction phase, the return cruise, and the Earth return phase. The complexity, mass, and power requirements of this system, or combination of systems, will depend on the details of these mission phases, the final sample temperature chosen, and the sample volume. Passive cooling of the sample return container will be a key component to the cryogenic system during all mission phases but will likely be the primary system during the cruise phase.  A cryocooler system can provide the auxiliary cooling needed in all mission phases but will be especially important during the Earth return phase, regardless of mode (section \ref{section:returnmode}). The temperature achievable with a passive cooling system or a cryocooler depends heavily on the parasitic heat loss in the system,  which  scales roughly with the geometric size of the cryostat. The sample size returned and the sample storage temperature required will dictate the size and complexity of the cryogenic system. 60\,K sample temperatures are readily achieved with current systems with extensive flight experience.

{A trade study should be undertaken of mass and power requirements for the cryogenic systems as a function of sample temperature,  sample volume, and Earth return mode, including assessment of the science that can be done as a function of the cryogenic temperature of the sample.}  The return mode (section \ref{section:returnmode}) in particular will strongly influence cryocooler system mass and shock and vibration requirements.

\section{Direct Earth return vs. space-based recovery}
\label{section:returnmode}

\citet{veverka2008}  found that the requirement for maintenance of cryogenic temperatures from sampling through curation had the strongest mission design implications for post-re-entry recovery. The possibility of a long delay before recovery implied the requirement for up to  hundreds of pounds of batteries  to power the cryocooler on the ground, depending on the sample volume and thermal requirements. The possibility of a hard landing would also drive a mechanically robust (hence heavy and expensive) cryocooler design. Phase change materials could be also be used to complement the cryocooler, depending on the temperature requirements.    Finally, as \citet{veverka2008} noted, a hard landing, even under a parachute, could compromise the stratigraphic integrity of a core sample.  

{A trade study on mission architectures for two return modes should be undertaken:   a space-based return, perhaps enabled by the ISS, the Deep Space Gateway (DSG) or the SpaceX Starship, versus a Stardust/Genesis style direct Earth return.} Logistical and safety issues in interfacing with the human spaceflight program will be significant, although if a phase change buffer is not used, the safety issues are not obviously more severe than those already encountered on previous missions (e.g., Trek, a fragile glass instrument returned from {\em Mir} by the US space shuttle \citep{Westphal1998}).  Cryocooler He coolant volumes are sufficiently small that even if completely vented into a crew cabin, the He concentration would still be $\ll$1\%.   If these challenges can be overcome, the advantages of space-based return could be substantial: {\bf 1:} Once retrieved, robust power for the cryocooler could be supplied by DSG or crew capsule power. {\bf 2:} Shock and vibration limits imposed by crew would be much less severe, reducing the risk of compromising the stratigraphy {\bf 3:}
Because of safety requirements for human spaceflight, the likelihood of a Genesis-style mishap will be significantly reduced

\section{Sample recovery, transport, curation and analysis}

\label{section:handling}


A cryogenic sample returned directly to Earth (probably at Utah Test and Training Range, UTTR) would require  recovery and placement into a ground-based cryocooler for transport to a UTTR cleanroom.  This would require tracking of the re-entering sample return capsule (SRC) to the ground, which is a well-utilized capability of UTTR, and helicopter support. There is heritage for most of these operations, in a non-cryogenic mode, from the recovery of Genesis and Stardust return capsules at UTTR.  A cryocooler containing the SRC would be flown to Houston and the JSC Curation Facility where the sample container would be removed from the SRC and secured in permanent cryogenic curation facilities.  

A cryogenic sample returned via space-based recovery would follow a similar trajectory, except that the sample might stay in the flight cryocooler, which would be externally powered, through delivery to JSC curation, and the samples would be unloaded from the cryocooler in the curation facility. This would reduce risk significantly by eliminating some sample-handling steps in the field. 

\label{section:curation}

In the astromaterials community, handling and curation of frozen samples has been limited to some of the Tagish Lake meteorite samples \citep{Zolensky2002,Herd2016}, and some Apollo lunar samples, recently made available for study through the ANGSA program, that have been stored and processed at 251\,K. Fortunately, NASA has recognized the inevitability of cryogenic sample curation and handling, and thus in 2018 began construction of the necessary advanced curation laboratories at the JSC Curation Facility \citep{McCubbin2020}. These critical capabilities will have been developed, practiced, and be well understood by the time samples are returned to Earth.   

\label{section:analysis}

Some key attributes of the cometary building blocks can only be analyzed by coordinated microanalysis of cryogenically returned samples. Understanding the spatial relationships among the various refractory and volatile components is essential to expanding our understanding of the origins of the individual cometary components, and the extent to which those components evolved and interacted through time. These detailed analyses can only be performed in ground-based labs.

Laboratory instrumentation for cryogenic sample handling and analysis has advanced dramatically over the last decade. The advent of cryomicroscopy of biological samples has led to a profusion of commercially available tools for  focused ion beam (FIB) microsampling, cryotransfer, and cryoelectron microscopy at 10s of K.  The viability of isotopic measurements of meteoritical samples held at cryogenic temperatures has been demonstrated \citep{Yurimoto2014}. Cryogenic (77\,K) stages  are currently available at several synchrotron x-ray beamlines. The Department of Energy recently identified the development of electron microscopes and samples stages that enable work at 5\,K with $<$0.1\,nm resolution for non-biological samples  as a priority for addressing multiple agency Grand Challenges \citep{DOEreport}.  These methods and instruments are already on the path to widespread adoption across the microanalysis research sector, and should be integrated into the Decadal Survey as part of the NASA technology roadmap for sample return.


\setlength{\bibsep}{0.0pt}

\clearpage

\end{document}